\documentclass[letterpaper,twocolumn,prl,aps,superscriptaddress,amsmath,amssymb,floatfix]{revtex4-1}
\usepackage{mathptmx}
\usepackage[latin9]{inputenc}
\usepackage{color}
\usepackage{amsmath}
\usepackage{amssymb}
\usepackage{graphicx}
\usepackage{esint}
\usepackage[unicode=true,
 bookmarks=true,bookmarksnumbered=false,bookmarksopen=false,
 breaklinks=false,pdfborder={0 0 1},backref=false,colorlinks=true]
 {hyperref}
\hypersetup{
 linkcolor=magenta,urlcolor=blue,citecolor=blue,pdfstartview={FitH},hyperfootnotes=false}

\makeatletter

\usepackage{textcomp}
\usepackage{epstopdf}

\usepackage{amsfonts}

\pdfpageheight\paperheight
\pdfpagewidth\paperwidth

\@ifundefined{textcolor}{}{%
 \definecolor{BLACK}{gray}{0}
 \definecolor{WHITE}{gray}{1}
 \definecolor{RED}{rgb}{1,0,0}
 \definecolor{GREEN}{rgb}{0,1,0}
 \definecolor{BLUE}{rgb}{0,0,1}
 \definecolor{CYAN}{cmyk}{1,0,0,0}
 \definecolor{MAGENTA}{cmyk}{0,1,0,0}
 \definecolor{YELLOW}{cmyk}{0,0,1,0}
}

\usepackage{xcolor}\usepackage{soul}
\newcommand{\bra}[1]{\ensuremath{\left\langle#1\right|}}
\newcommand{\ket}[1]{\ensuremath{\left|#1\right\rangle}}

\definecolor{blue}{rgb}{0,0,1}
\definecolor{red}{rgb}{1,0,0}
\definecolor{green}{rgb}{0,1,0}

\usepackage{soul}

\makeatother

\begin{document}

\affiliation{Laboratory of Quantum Information, University of Science and Technology of China, Hefei 230026, China}
\affiliation{Center for Quantum Information, Institute for Interdisciplinary Information Sciences, Tsinghua University, Beijing, China}
\affiliation{Anhui Province Key Laboratory of Quantum Network,
University of Science and Technology of China, Hefei 230026, China}
\affiliation{CAS Center For Excellence in Quantum Information and Quantum Physics, University of Science and Technology of China, Hefei, Anhui 230026,
China}
\affiliation{Hefei National Laboratory, Hefei 230088, China}

\title{Scalable Generation of Macroscopic Fock States Exceeding 10,000 Photons}

\author{Ming Li}
\affiliation{Laboratory of Quantum Information,
University of Science and Technology of China, Hefei 230026, China}
\affiliation{Anhui Province Key Laboratory of Quantum Network,
University of Science and Technology of China, Hefei 230026, China}
\affiliation{Hefei National Laboratory, Hefei 230088, China}

\author{Weizhou Cai}
\affiliation{Laboratory of Quantum Information,
University of Science and Technology of China, Hefei 230026, China}
\affiliation{Anhui Province Key Laboratory of Quantum Network,
University of Science and Technology of China, Hefei 230026, China}

\author{Ziyue Hua}
\affiliation{Center for Quantum Information, Institute for Interdisciplinary Information
Sciences, Tsinghua University, Beijing, China}

\author{Yifang Xu}
\affiliation{Center for Quantum Information, Institute for Interdisciplinary Information
Sciences, Tsinghua University, Beijing, China}

\author{Yilong Zhou}
\affiliation{Center for Quantum Information, Institute for Interdisciplinary Information
Sciences, Tsinghua University, Beijing, China}

\author{Zi-Jie Chen}
\affiliation{Laboratory of Quantum Information,
University of Science and Technology of China, Hefei 230026, China}
\affiliation{Anhui Province Key Laboratory of Quantum Network,
University of Science and Technology of China, Hefei 230026, China}

\author{Xu-Bo Zou}
\affiliation{Laboratory of Quantum Information,
University of Science and Technology of China, Hefei 230026, China}
\affiliation{Anhui Province Key Laboratory of Quantum Network,
University of Science and Technology of China, Hefei 230026, China}
\affiliation{Hefei National Laboratory, Hefei 230088, China}

\author{Guang-Can Guo}
\affiliation{Laboratory of Quantum Information,
University of Science and Technology of China, Hefei 230026, China}
\affiliation{Anhui Province Key Laboratory of Quantum Network,
University of Science and Technology of China, Hefei 230026, China}
\affiliation{Hefei National Laboratory, Hefei 230088, China}

\author{Luyan Sun}
\email{luyansun@tsinghua.edu.cn}
\affiliation{Center for Quantum Information, Institute for Interdisciplinary Information
Sciences, Tsinghua University, Beijing, China}
\affiliation{Hefei National Laboratory, Hefei 230088, China}

\author{Chang-Ling Zou}
\email{clzou321@ustc.edu.cn}
\affiliation{Laboratory of Quantum Information,
University of Science and Technology of China, Hefei 230026, China}

\affiliation{Anhui Province Key Laboratory of Quantum Network,
University of Science and Technology of China, Hefei 230026, China}
\affiliation{CAS Center For Excellence in Quantum Information and Quantum Physics,
University of Science and Technology of China, Hefei, Anhui 230026,
China}
\affiliation{Hefei National Laboratory, Hefei 230088, China}
%\date{\today}

\begin{abstract}
\textbf{The scalable preparation of bosonic quantum states with macroscopic excitations poses a fundamental challenge in quantum technologies, limited by control complexity and photon-loss rates that severely constrain prior theoretical and experimental efforts to merely dozens of excitations per mode. Here, based on the duality of the quantum state evolution in Fock state space and the optical wave-function propagation in a waveguide array, we introduce a Kerr-engineered multi-lens protocol in a single bosonic mode to deterministically generate Fock states exceeding $10,000$ photons. By optimizing phase and displacement operations across lens groups, our approach compensates for non-paraxial aberrations, achieving fidelities above $73\%$ in numerical simulations for photon numbers up to $N=100,000$. Counterintuitively, the protocol's execution time scales as $N^{-1/2}$ with the target photon number $N$, exhibiting robustness against the photon loss. Our framework enables exploration of quantum-to-classical transitions of giant Fock states, paving the way for advanced quantum metrology with significant quantum gains, and error-corrected quantum information processing in high-dimensional Hilbert spaces.}
\end{abstract}
\maketitle

\noindent \textbf{\large{}Introduction}{\large\par}
\noindent Fock states, defined as the quantum states of light containing an exact number of photons, reveal the fundamental non-classical nature of the electromagnetic field and have long stood as a cornerstone of quantum mechanics~\cite{Scully1997,Kok2010,Agarwal2012}. In recent decades, the preparation and manipulation of Fock states at the microscopic scale, involving only a few excitations, has been extended across diverse bosonic platforms, including mechanical oscillators~\cite{OConnell2010}, optical cavities~\cite{Gleyzes2007,Sayrin2011}, and superconducting microwave resonators~\cite{hofheinz2008generation,hofheinz2009synthesizing}.  These advances have enabled the construction of quantum optical technologies~\cite{Couteau2023,Flamini2019}, facilitated proof-of-principle demonstrations of quantum information processing, and deepened our understanding of quantum mechanics at its most fundamental level. Now, extending Fock state generation to hundreds or even thousands of photons becomes profoundly important, not only for exploring quantum phenomena at macroscopic scales~\cite{RevModPhys2003,RevModPhys2018}, but also for advancing quantum simulation~\cite{PRAfocklattice,wdwscience}, photonic quantum algorithms~\cite{wang2017high,Aaronson2011,zhongscience2020}, and quantum-enhanced sensing~\cite{Wolf2019,deng2024quantum}. Yet at macroscopic energy scales, quantum states of light remain restricted to thermal radiation or coherent states from laser sources, both exhibiting Gaussian statistics that follow semi-classical descriptions and obscure their underlying quantum nature.

\begin{figure*}
\begin{centering}
\includegraphics[width=1.0\textwidth]{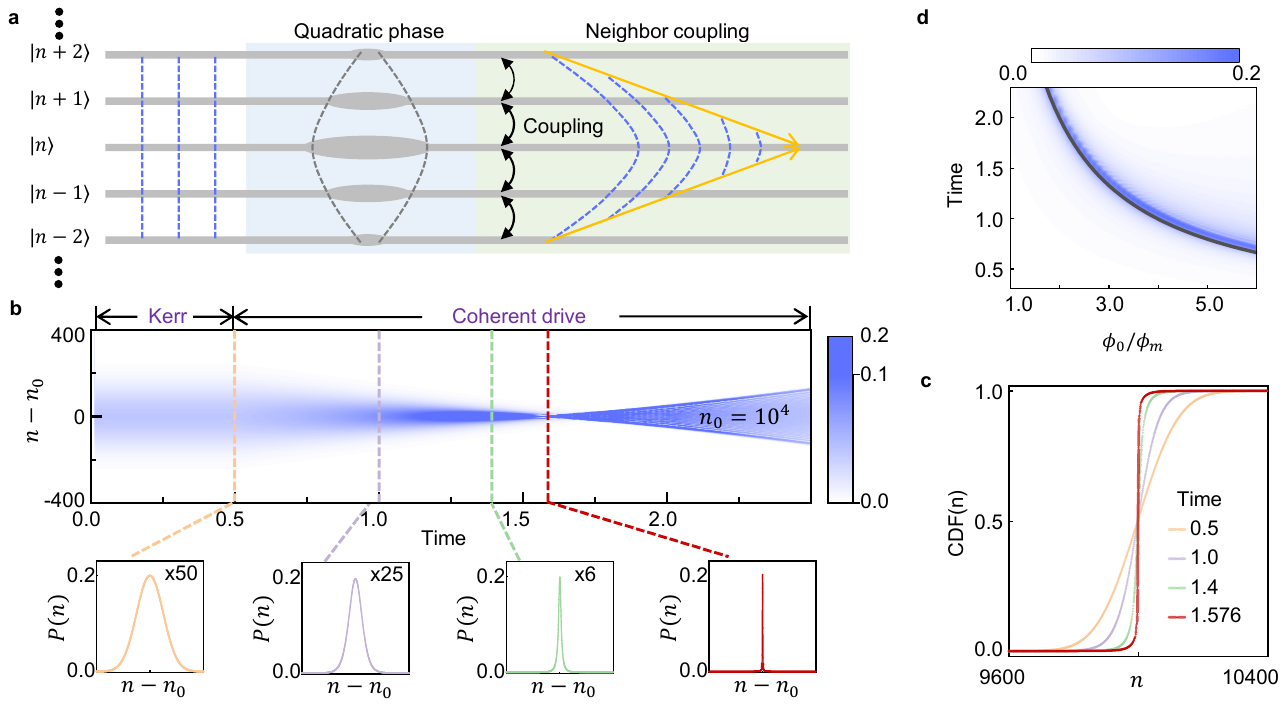}
\par\end{centering}
\caption{\textbf{Fock-space lens}. \textbf{a}, Illustration of Fock-space lens in the Fock-state lattice of a bosonic mode. Photon number population is squeezed by a quadratic phase accumulation followed by coherent coupling between neighborhood sites. \textbf{b}, Evolution of the photon number distribution $P(n,t)$. The coherent state $|\alpha\rangle$ with $\alpha=100$ accumulates a quadratic phase during $0\protect\leq t\protect\leq0.5$ and single-photon coherent drive is applied after $t=0.5$. Probability higher than $0.1$ is plotted as the same color for better illustration. Lower panel: the $P(n)$ at $t=0.5$, $t=1.0$,  $t=1.4$ are multiplied by 50, 25, and 6, respectively.  \textbf{c}, The accumulative distribution function $\text{CDF}(n)$ of $P(n)$ at different times.
\textbf{d}, Maximum of $|c_{n}|^{2}$ for different phase $\phi_0$ and single-photon drive time $t$. The phase is normalized to $\phi_{m}=2.45\times 10^{-3}$. The strength of coherent drive is $\varepsilon_{p}=1$. The condition for maximal $|c_{n}|^{2}$ coincides well with the analytical expression between focal time and phase shown in Eq.\,($\ref{eq:focal_lattice}$).}
\label{Fig1}
\end{figure*}

However, advancing Fock state generation into the macroscopic regime presents formidable challenges. Existing strategies include probabilistic measurement-based post-selection~\cite{PhysRevA2019Fock,deng2024quantum} and deterministic schemes implemented in optical cavity quantum electrodynamics (QED)~\cite{sayrin2011real,deleglise2008reconstruction,PhysRevLett.108.243602,Cosacchi2020,cqed}, optical waveguide QED~\cite{PhysRevLett.118.213601,PhysRevLett.115.163603}, and superconducting circuits~\cite{hofheinz2009synthesizing,PhysRevLett.104.100504,Chu2018}. These approaches typically rely on sequential photon-number transitions~\cite{hofheinz2008generation,PhysRevLett.101.240401,premaratne2017microwave,fluoquet}, global quantum control~\cite{heeres2017implementing,cqed}, or reservoir engineering~\cite{Rivera2023}. Yet, they all confront severe scalability bottlenecks. The operational complexity and required resources scale unfavorably with the target photon number $N$. More critically, the fundamental limitation imposed by inevitable environmental coupling leads to an effective decay rate that scales linearly with the photon number~\cite{PhysRevLett.101.240401}. Consequently, the lifetimes of macroscopic Fock states diminish drastically as $N$ increases. The absence of scalable preparation protocols, combined with the intrinsic trade-off between increasing control complexity and decreasing state lifetime, severely limits the achievable Fock state size, leaving state-of-the-art demonstrations confined to the order of $\sim 100$ photons~\cite{deng2024quantum,cqed}.

Here, we propose a general and scalable scheme for the generation of Fock states with giant photon numbers. By utilizing simple coherent drive and Kerr nonlinearity of a bosonic mode, we construct effective lenses in Fock space, which precisely focus the photon-number distribution, transforming an initial state into a target high-excitation state. Our method exhibits a unique two-fold scalability. First, the control complexity does not increase with the photon number, fundamentally overcoming the scalability bottleneck of existing methods. Second, the required preparation time scales as $\sim 1/\sqrt{n}$. This accelerated dynamics effectively mitigates the impact of the linearly growing environmental decay rate, demonstrating strong robustness against loss. Our work provides a novel perspective for preparing and manipulating quantum states with macroscopic energy, opening new avenues for exploring quantum physics and practical quantum metrology in the large-photon-number regime.

\smallskip{}
\noindent \textbf{\large{}Results}{\large\par}
\noindent
Figure~\ref{Fig1}a illustrates the correspondence between optical wave propagation in a waveguide array and quantum state evolution in a bosonic mode. The optical wave-function $|\psi\rangle=\sum_n c_n|n\rangle$ in the waveguide array, with the basis $\ket{n}$ denoting the individual waveguide, propagates and evolves in the array via nearest-neighboring hopping. Such wave propagation resembles the optical beam propagation in a discretized space. Consequently, the wave-function across a large array of waveguides can be efficiently and precisely manipulated through the well-established principles of optics, which have been extensively studied experimentally~\cite{Garanovich2012}. When a quadratic phase modulation is introduced across the waveguides, mimicking a lens in free-space optics, the wave-function spreading in the array could be focused.

\begin{figure*}
\begin{centering}
\includegraphics[width=1.0\textwidth]{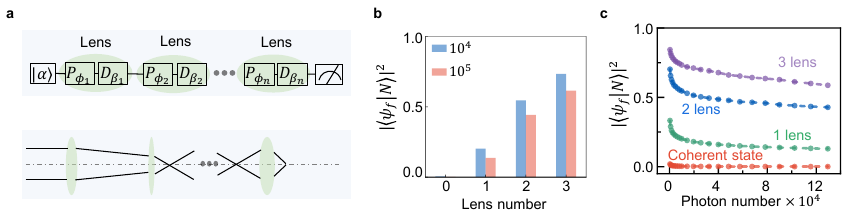}
\par\end{centering}
\caption{\textbf{Fock state generation using a group of Fock-space lenses}. \textbf{a}, A series of lens operations are applied to the initial coherent state (upper panel), resembling a lens group for tight focusing of a light beam (lower panel). By optimizing the phase and displacement of each lens, the fidelity of the target Fock state can be greatly improved. \textbf{b}, Optimized fidelity to obtain a Fock state with photon numbers $N=10,000$ and $N=100,000$ for different number of lenses. Using three Fock-space lenses, the state fidelity reaches $73.3\,\%$ and  $61.5\,\%$ for $10,000$- and $100,000$-photon Fock states, respectively. \textbf{c}, Relationship between the optimized fidelity and the target Fock state for different lens-group configurations.
}
\label{Fig2}
\end{figure*}

The duality can be established by treating each waveguide site as a distinct Fock state $\ket{n}$ of a bosonic mode, so that the corresponding quantum state stored in the mode can be represented by the state $|\psi\rangle=\sum_n c_n|n\rangle$. For a general model of a bosonic mode with annihilation operator $a=\sum_n\sqrt{n}\ket{n-1}\bra{n}$, a coherent drive
\begin{equation}
    \varepsilon_{p}(a^\dagger+a)=\varepsilon_{p}\sum_n\sqrt{n}(\ket{n-1}\bra{n}+\ket{n}\bra{n-1})
\end{equation}
simulates coherent hopping between neighboring Fock states, while the Kerr nonlinearity
\begin{equation}
    -\chi a^{\dagger}a^{\dagger}aa=-\chi \sum_n n(n-1)\ket{n}\bra{n}
\end{equation}
imparts a quadratic phase across the Fock-state lattice. Here, $\chi$ is the Kerr coefficient and $\varepsilon_{p}$ is the single-photon drive strength. Based on the duality, the focusing of a quantum state spreading in Fock-state lattice to a narrow distribution can be realized by switching the Kerr coefficient and the coherent drive, thereby implementing the quadratic phases and the propagation evolution of the state. We term such operation on the Fock-state lattice as a Fock-space lens. The focal center of the lens can be tuned to an arbitrary photon number $n_0 = (\chi-\Delta)/2\chi$ by adjusting the drive detuning $\Delta$.

Figure$\,$\ref{Fig1}b shows the evolution of the photon number distribution $P(n,t)=|c_{n}(t)|^{2}$ of the Fock space lens operation, for an initial state of the mode prepared to a coherent state $|\psi\rangle=|\alpha\rangle$ with a mean photon number $|\alpha|^{2}=10^{4}$. During time $0\leq t\leq0.5$, the quantum state accumulates a photon number-dependent phase $\phi(n)=\phi_{0}(n-|\alpha|^{2})^2$ under Kerr nonlinearity, with $\phi_{0}=\chi \tau_{\mathrm{Kerr}}$ determined by the Kerr interaction duration $\tau_{\mathrm{Kerr}}$. At $t=0.5$, the single-photon drive $\varepsilon_{p}\left(a+a^{\dagger}\right)$ is applied to couple different Fock states and the quantum state starts to focus. We find that the distribution $P(n,t)$ becomes progressively narrower for $t\leq1.576$, resembling the phenomenon of beam focusing after a lens. The photon number distributions at different instants are plotted in the lower panels of Fig.$\,$\ref{Fig1}b. Compared to the initial coherent state {[orange]} that occupies hundreds of Fock states, only a few Fock states have a significant population at the focal spot {[red]}. The peak value of $P(n)$ increases from $0.004$ to $0.2$, demonstrating the powerful focusing ability of the Fock-space lens. The width of photon number distribution can be seen more explicitly by the cumulative distribution function $\text{CDF}(n)$ of $P(n)$, where a very sharp step is observed at the focal spot {[Fig.\,\ref{Fig1}c]}. Different from the continuous wave equation in optics, the wave-function in $n$-distribution after the focal point is a superposition of several Bessel functions that is strongly correlated to Fock state orders, due to the discrete nature of Fock lattice~\cite{Peschel1998}.

Analogous to the relationship between focal length and lens curvature in optics, the required interaction for focusing the states, i.e., the drive duration $\tau_{\mathrm{drive}}$, scales inversely to the thickness of the lens ($\tau_{\mathrm{Kerr}}$), i.e., the quadratic phase coefficient $\phi_{0}$, given by
\begin{equation}
\tau_{\mathrm{drive}}  =  \frac{1}{4\sqrt{n_0}\varepsilon_p\phi_{0}}.
\label{eq:focal_lattice}
\end{equation}
Figure$\,$\ref{Fig1}d shows the achievable optimal Fock state probabilistic $P(n)$ for different quadratic phase coefficient $\phi_{0}$ and drive interactions. The blue stripe that corresponds to the working parameters of the focal point shows a negative relationship between the optimal value of drive time  $\tau_{\mathrm{drive}}$ and $\phi_0$, which coincides well with the analytical derivations according to optical principles [black line].

%The applicability of Fock-space lenses is subject to two limitations during practical implementation. First,
%the coupling strength between different Fock-state sites under coherent drive exhibits a $\sqrt{n}$-scaling with respect to the photon number for a realistic bosonic mode. The
%non-ideal diffusion equation compromises
%the performance of the Fock-space lens for quantum states with small photon numbers. To implement an ideal Fock-space lens in the low photon number region, advanced control techniques~\cite{baker2021heisenberg} as well as novel superconducting circuit~\cite{liu2021proposal} can be designed to realize Susskind-Glogower bare operator~\cite{bareoperator} to eliminate the $\sqrt{n}$ factor.
%While for large photon numbers as in Fig.~\ref{Fig1}, the influence of $\sqrt{n}$-scaling is negligible.
%Second, the implement of Fock-space lens requires a quantum state to have a narrow $k$-distribution
%to ensure a reasonable truncation of $\cos(k)$ to second-order $k^{2}$, which is not satisfied for quantum states with broad $k$-distribution, for example, an ideal Fock state.
%For initial coherent states, its $k$-distribution is greatly broadened by the quadratic phase.
%The invalidity of the truncation approximation manifests as non-paraxial behavior or aberration in optical beam propagation dynamics.

\begin{figure*}
\begin{centering}
\includegraphics[width=1.0\textwidth]{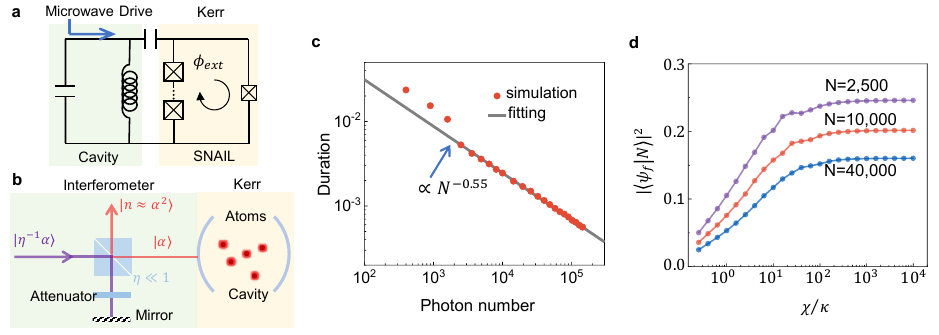}
\par\end{centering}
\caption{\textbf{Experimental feasibility}. \textbf{a}, Microwave realization: a superconducting circuit comprising a capacitively-coupled microwave cavity and a SNAIL. The SNAIL provides the Kerr nonlinearity required for quadratic phase accumulation. \textbf{b}, Optical realization: a strong laser beam passes through a beam splitter with transmittance $\eta \ll 1$ and obtain the parabolic phase from a cavity-QED system. The reflected field is then displaced by the reflected laser from the downside mirror.  \textbf{c}, The normalized time duration of the Fock lens, defined as the ratio of the optimal second-order coefficient $\phi_{0}$ to the Kerr coefficient $\chi$. Dots: simulation. Line: power-law fit for photon numbers $> 2500$, yielding a decay exponent of $-0.5525$. \textbf{b}, Relationship between the fidelity and the Kerr-to-loss ratio $\chi/\kappa$ for Fock states of $2500$, $10000$, and $40000$ photons.}

\label{Fig3}
\end{figure*}

Although the duality between the optical and Fock-space lenses enables significant narrowing of photon-number distributions, imperfections, such as the photon-number-dependent ($\propto\sqrt{n}$) hopping terms, manifest as non-paraxial behavior or aberration in optical beam propagation. These aberrations become the main obstacle for a single lens to generating high-fidelity Fock states with giant photon numbers. However, such aberrations can be mitigated by employing a group of Fock-space lenses, similar to the lens group widely adopted in optics field {[}Fig.$\,$\ref{Fig2}a{]}. By repetitively applying quadratic phase imprinting operation $P_{\phi}$ and displacement operation $D(\beta)$ on an initial coherent state, the fidelity between the focused state and target Fock state can be substantially improved.
Due to the non-commuting nature between Kerr nonlinearity and displacement operator, high-order phase corrections can be introduced to compensate the aberration, thus enhancing the focusing capability of the lenses. By optimizing the quadratic phase coefficients $\phi_j$ of each phase operation and the strength $\beta_j$ of each displacement, we calculate the maximal achivable fidelity $F=|\langle \psi|N\rangle|^2$ between the focused state $|\psi\rangle$ and a target Fock state $|N\rangle$, for a coherent state input. As shown in Fig.~\ref{Fig2}b, the fidelity for generating a Fock state with $N=10,000$ photons can be improved to exceed $0.55$ with two lenses and to exceed $0.73$ with three lenses.

The discrete nature of the Fock lattice causes deviations from the continuous diffusion equation for ideal lensing, leading to reduced fidelity as photon number increases, as demonstrated in Fig.$\,$\ref{Fig2}c. Remarkably, however, the fidelity does not decrease significantly with the increase of the photon number. For the case of three Fock lenses, the fidelity still maintains above $0.61$ even for $100,000$-photon Fock state {[}Fig.$\,$\ref{Fig2}b{]}. Fitting the dependence of the fidelity $F$ on $N$ to a power low $F=xN^{-y}$, we extract a decay exponent $y_{0}=0.5$ for a bare coherent state without lens, yielding the well-known $\sqrt{N}$-scaling of the Poisson statistics. In contrast, the fitted decay exponents are $y_{1}=-0.159$, $y_{2}=-0.080$ and $y_{3}=-0.047$ for 1, 2 and 3 lenses, respectively. These near-zero decay exponents indicate a highly scalable route to macroscopic Fock state with photon number comparable to the energy of lasers in the classical world, which is potential for investigating macroscopic quantum phenomena and practical application of quantum metrology.

Bosonic mode with coherent drive and controllable Kerr interaction can be realized in various quantum systems. Figure~\ref{Fig3}a illustrates a superconducting microwave cavity coupled to a SNAIL via a capacitor. The strong Kerr nonlinearity of the SNAIL that is controlled by the external flux $\phi_{\text{ext}}$ can be transferred to the cavity via the capacitive coupling, and the detuning is flexibly
tuned by the drive frequency. Then the phase operation and displacement of Fock-space lens are implemented in the cavity by switching the external flux and microwave drive. Figure~\ref{Fig3}b illustrates the preparation of giant Fock state in the optical band. A strong laser beam described by coherent state $|\eta^{-1}\alpha\rangle$ passes through a beam splitter with transmittance $\eta \ll 1$ and interacts with a cavity-QED system. The cavity-QED system consisting of multi-level atoms induces an effective Kerr effect on the optical cavity and adds a quadratic phase on the reflected light. The reflected field then mixes with the incident laser and acquires a displacement operation and the strength of displacement is controlled by attenuator. Fock state $|N\approx \alpha^2\rangle$ is finally obtained in the upper port of the interferometer.

When operation on quantum state with thousands of photons, photon
losses inevitably degrade the fidelity of the final state. For a target $N$-photon Fock state, its decay rate is enlarged
by $N$ times to $N\kappa$, with $\kappa$ being the photon dissipation
rate of the bosonic mode.
This poses strict requirements on the execution time of the Fock-space lens for a high fidelity Fock state. In practice, the initial
coherent state can be easily prepared to arbitrary large value by
simple coherent drives, which is immune to the single-photon loss
operator $a$. During the execution of the Fock-space lens, the single-photon
drive costs time $\tau_{\mathrm{drive}}$ {[}Eq.$\,($\ref{eq:focal_lattice}){]}, which can
be minimized by increasing the driving strength $\varepsilon_{p}$.
However, the execution time of phase operation is inverse to the strength $\chi$ of Kerr nonlinearity. To reduce the influence of dissipation during the
phase operation process, it is beneficial to have a physically inaccessible large Kerr-to-loss ratio $\chi/\kappa$ or a small quadratic phase coefficient $\phi_{0}$ to shorten the phase operation time.

Fortunately, the optimal phase coefficient $\phi_{0}$ decreases with increasing photon number in the Fock-lens scheme.
Considering an initial coherent states $|\alpha\rangle$, the wave-function in the $k$-space approximates a Gaussian function $C_{k}\propto e^{-\frac{\alpha^2}{4}k^2}$, which transforms to a broader Gaussian function $e^{-\frac{k^2}{4\alpha^2\phi_{0}^2}}$ after the phase operation. A high-fidelity focused state requires $\alpha\phi_{0}$ as large as possible to match the $k$-distribution of Fock states, however is limited by the paraxial condition $\alpha\phi_{0}<1$. The maximum phase $\phi_{0}$ for Fock state $|n\rangle$ thus scales as $\phi_{0}\propto\alpha^{-1}=N^{-1/2}$, which also leads to a decreasing phase operation time $\tau_{\mathrm{Kerr}}\propto N^{-1/2}$. Figure$\,$\ref{Fig3}(a) shows the normalized time duration of Fock lens (optimal phase coefficient $\phi_{0}$)
to maximize the fidelity of Fock state $|n\rangle$ from a single lens.
Applying the power function fitting $\phi_{0}=xn^{-y}$ to the numerical data for photon number above $2,500$ {[}dots,
Fig.$\,$\ref{Fig3}(c){]} gives $y=-0.55$, shown by the dashed line. It indicates that for a given Kerr nonlinearity, the execution time of the Fock lens decreases by $N^{-0.55}$, which matches the predicted value $-1/2$ in the above analysis. This decreasing execution time shows stark contrast to conventional methods, whose quantum control complexity and time both increase with the photon number.

The decreasing execution time makes the lens highly robust to single-photon
loss. By simulating the open system using the master equation, we investigate
the performance of the lens under different Kerr-to-loss ratio. Figure$\,$\ref{Fig3}d
shows the relationship between the fidelity and Kerr-to-loss ratio $\chi/\kappa$ for Fock
states with $2,500$, $10,000$, and $40,000$ photons based on one Fock-space lens. The fidelity
increases with $\chi/\kappa$ for all Fock states and approaches their ideal values for $\chi/\kappa>100$. Even for $\chi/\kappa\sim1$, the Fock-space lens still maintains remarkable performance, as
the fidelity is increased  by nearly $6$ times after the lens operation.

\smallskip{}
\noindent \textbf{\large{}Discussion}{\large\par}
\noindent  A simple scheme, which requires only coherent drive and Kerr nonlinearity of a bosonic mode, is proposed to generate Fock states with giant photon numbers. The scheme exhibits excellent scalability as both the control complexity and the preparation time do not increase with the photon number. Using this approach, we numerically demonstrate the preparation of Fock states beyond $10,000$ photons with a high fidelity above $73\,\%$. The approach is feasible for various experimental quantum systems, including superconducting circuits, ion traps, cavity-QED, and acoustic systems. Such states enter a truly macroscopic quantum regime, as the equivalent circulating optical power in a cavity QED implementation~\cite{Yang2023} reaches 4\,mW when $N>10,000$. The state fidelity can be further improved toward near-unity by applying sequential parity measurements and cavity multiplexing. Moreover, for quantum metrology applications, the near-zero decay exponent of fidelity allows for metrology gain approaching the Heisenberg limit.
The mechanism to manipulate the quantum states of photons in the quantizied photon-number space is a direct analog of manipulating classical wave optics in the time and space dimensions~\cite{Kolner1994,Karpinski2017}. This connection will excite new quantum state manipulation methods by drawing analogies to spatial optical components and systems. In the large photon number regime, new quantum optics phenomena, such as the transition between quantum and classical optics can be explored, thereby advancing both our understanding of macroscopic quantum physics and the development of practical quantum technologies.

\smallskip{}

\begin{acknowledgments}
This work was funded by the National Natural Science Foundation of China (Grants No. 92265210, 92365301, 92165209, 92565301, 12550006, 12547179 and 12574539) and the Quantum Science and Technology-National Science and Technology Major Project (2021ZD0300200). This work is also supported by the Fundamental Research Funds for the Central Universities, the USTC Research Funds of the Double First-Class Initiative, the supercomputing system in the Supercomputing Center of USTC the USTC Center for Micro and Nanoscale Research and Fabrication.
\end{acknowledgments}

\end{document}